\documentclass{svproc}
\usepackage{url}
\usepackage{graphicx}        
\usepackage{amsfonts}
\usepackage{cite}

\begin{document}
\mainmatter              
\title{Investigating Centrality Measures in\\ Social Networks with Community Structure}
\titlerunning{Investigating Centrality Measures}  
%
\author{Stephany Rajeh\inst{1} \and Marinette Savonnet \inst{1} \and
	Eric Leclercq \inst{1} \and
	Hocine Cherifi \inst{1}}
\authorrunning{S. Rajeh et al.} 
%
\tocauthor{Ivar Ekeland, Roger Temam, Jeffrey Dean, David Grove,
	Craig Chambers, Kim B. Bruce, and Elisa Bertino}
\institute{Laboratoire d’Informatique de Bourgogne - University of Burgundy, Dijon, France\\
	\email{stephany.rajeh@u-bourgogne.fr}}

\maketitle    

\begin{abstract}
Centrality measures are crucial in quantifying the influence of the members of a social network. Although there has been a great deal of work dealing with this issue, the vast majority of classical centrality measures are agnostic of the community structure characterizing many social networks. Recent works have developed community-aware centrality measures that exploit features of the community structure information encountered in most real-world complex networks. In this paper, we investigate the interactions between 5 popular classical centrality measures and 5 community-aware centrality measures using 8 real-world online networks. Correlation as well as similarity measures between both type of centrality measures are computed. Results show that community-aware centrality measures can be divided into two groups. The first group, which includes Bridging centrality, Community Hub-Bridge and Participation Coefficient, provides distinctive node information as compared to classical centrality. This behavior is consistent across the networks. The second group which includes Community-based Mediator and Number of Neighboring Communities is characterized by more mixed results that vary across networks.
\end{abstract}

\section{Introduction}
With the rapid increase of online social networks (OSNs) such as Facebook and Twitter, large amount of data is being generated daily. A valuable mining area of network data is composed when OSNs are modeled into nodes and edges. Identifying key nodes in such networks is the basis of major applications such as viral marketing \cite{jalili2017information}, controlling epidemic spreading \cite{wang2017vaccination}, and determining sources of misinformation \cite{azzimonti2018social}.
Designing centrality measures is a main approach to quantify node influence. Numerous centrality measures exploiting various properties of the network topology have been developed \cite{lu2016vital}. Information exploited can be either in the neighborhood of the node or concerning all the topological structure of the network. The former called local centrality measures are less computationally expensive as compared to the later called global centrality measures. However, local centrality measures, usually, aren't as much as accurate as global centrality measures. Recent works tend to combine both local and global measures \cite{sciarra2018change,ibnoulouafi2018m}.
Real-world OSNs often exhibit a community structure in which groups of nodes are closely connected to each other and sparsely connected to nodes in other communities \cite{girvan2002community, jebabli2015user}. Community structure has major implications on the dynamics of the network \cite{cherifi2019community}. To this end, researchers have taken classical centrality measures a step further to incorporate community structure information \cite{hwang2006bridging, ghalmane2019immunization, guimera2005functional, tulu2018identifying,gupta2015community, chakraborty2016immunization, kumar2018efficient, ghalmane2019centrality}. Community-aware centrality measures can be divided into two groups. The former explicitly rely on the community structure. They incorporate information about the type of links in a community (intra-community links and inter-community links). The latter targets ``bridges" that lie between communities without extracting the community structure information.

As classical centrality measures neglect the community structure, this raises a key question. Do community-aware centrality measures provide distinctive information about the members within OSNs when compared to classical centrality measures? Previous works have studied the relationship between classical centrality measures \cite{li2015correlation, oldham2019consistency, shao2018rank, Landherr2010, grando2016analysis} and between classical and hierarchy measures \cite{rajeh2020interplay}.
Nonetheless, to our knowledge, there is no previous work on the relationship between classical and community-aware centrality measures on OSNs. To fill this gap, here, 5 classical and 5 community-aware centrality measures are used in a comparative evaluation involving  8 real-world OSN. The community structure of the networks is extracted using the Infomap \cite{rosvall2008maps} community detection algorithm. Then, Kendall's Tau correlation and RBO similarity are calculated on all the possible combinations between the classical and community-aware centrality measures. Two groups of community-aware centrality measures can be seen. The first group provides distinctive information when compared against classical centrality measures and is consistent across the networks under study. It includes Bridging centrality, Community Hub-Bridge, and Participation Coefficient. The second group shows varying correlation and similarity on networks. It includes Community-based Mediator and Number of Neighboring Communities.

The paper is organized as follows. Classical and community-aware centrality measures alongside basic definitions are provided in section \ref{sec:Def}. The datasets and tools are provided in section \ref{sec:mm}. Experimental results are discussed in section \ref{sec:res}. Finally, the conclusion and future works are provided in section \ref{sec:conc}.

\section{Preliminaries and Definitions}
\label{sec:Def}
In this section preliminaries and definitions used throughout the rest of the paper are given.

\begin{itemize}
\item Consider a undirected and unweighted OSN as $G(V,E)$ where $V$ is the set of nodes
and $E \subseteq V \times V$ is the set of edges and $N=|V|$ is the total size of the network.
Nodes represent individuals and edges represent
social links between these individuals. The semantics of the social links depend on the platform of the OSN.
\item Consider $A = (a_{i,j})$ as the adjacency matrix showing connectivity of the network
$G$ such that $a_{i,j}= 1$, if node $i$ is connected to node $j$ and $a_{i,j}= 0$, otherwise.
\item Let the neighborhood of any node $i$ be defined as the set $\mathcal{N}_p(i)={\{j \in V:
(i,j) \in E\}}$ at length $p$, where $p=1,2,...,D$. $D$ is the diameter of $G$. Accordingly, two
nodes are neighbors of order $A^p$ if there's a minimal path connecting them at $p$ steps.
\item Let $C$ be the set of communities $C=\{c_1, c_2, ..., c_k\}$. The intra-community links are obtained from the graph $G_l$ where all inter-community links of the nodes are removed.
The inter-community links are obtained from the graph $G_g$ where all intra-community links of the nodes are removed.
\end{itemize}

\subsection{Classical Centrality Measures}
Following are the definitions of the 5 most popular centrality measures used in the study.

\subsubsection{Degree Centrality} is simply the total number connections a node has in the network. It is defined as follows:
\begin{equation}
\alpha_{d}(i)=\sum_{j=1}^{N}a_{ij}
\end{equation}
where $a_{ij}$ is obtained from $A^{1}$, 1-step neighborhood ($p$=1).

\subsubsection{Betweenness Centrality} captures the number of times a node falls between the shortest paths linking other node pairs. It is defined as follows:
\begin{equation}
\alpha_b(i)=\sum_{s,t\neq i }{\frac{\sigma_{i}(s,t)}{\sigma(s,t)} }
\end{equation}
where $\sigma(s,t)$ is the number of shortest paths between nodes $s$ and $t$ and $\sigma_{i}(s,t)$ is the number of shortest paths between nodes $s$ and $t$ that pass through node $i$.

\subsubsection{Closeness Centrality} is inverse of sum of geodesic distances to every other node from a given node. It is defined as follows:
\begin{equation}
\alpha_c(i)=\frac{N-1}{\sum_{j=1}^{N-1}d(i,j)}
\end{equation}
where $d(i,j)$ is the shortest-path distance between node $i$ and $j$.

\subsubsection{Katz Centrality} is based on how many nodes a node is connected to and also to the connectivity of its neighbors . It is defined as follows:
\begin{equation}
\alpha_k(i)= \sum_{p=1} \sum_{j=1} s^p a^p_{ij}
\end{equation}

where $a^p_{ij}$ is the connectivity of node $i$ with respect to all the other nodes at $A^p$ and $s^p$ is the attenuation factor where $s \in$ [0,1].

\subsubsection{PageRank Centrality} quantifies a node's importance similarly to Katz centrality with an additional layer based on a random surfer. It is defined as follows:
\begin{equation}
\alpha_p(i)=\frac{1-d}{N} + d \sum_{j \in \mathcal{N}_1(i)} \frac{\alpha_p(j)}{k_j}
\end{equation}
where $\alpha_p(i)$ and $\alpha_p(j)$ are the PageRank centralities of node $i$ and node $j$, respectively, $\mathcal{N}_1(i)$ is the set of direct neighbors of node $i$, $k_j$ is the number of links from node $j$ to node $i$, and $d$ is the damping parameter where $d \in$ [0,1], set to 0.85 in the experiments.

\subsection{Community-aware Centrality Measures}
Following are the definitions of the 5 community-aware measures of centrality used:
\subsubsection{Number of Neighboring Communities (NNC) \cite{ghalmane2019immunization}} is based on the number of communities a node can reach in one hop. For a node in community $c_k \subset C$, it is defined as follows:
\begin{equation}
\beta_{NNC}(i) = \sum_{c_l \subset C \backslash c_k}^{N} \bigvee_{j \in c_l} a_{ij}
\end{equation}
where $\bigvee_{j \in c_l} a_{ij} = 1$ when node $i$ is connected to at least one node $j$ in community $c_l$.

\subsubsection{Community Hub-Bridge (CHB) \cite{ghalmane2019immunization}} assumes a node simultaneously can act as a hub and a bridge. It combines the intra-community and inter- community links by weighting the former with the community size and the latter with the number of neighboring communities. For a node in community $c_k \subset C$, it is defined as follows:
\begin{equation}
\beta_{CHB}(i) = h_i(c_k) + b_i(c_k)
\end{equation}
where hub influence is given by $h_i(c_k)= |c_k| \times k_i^{intra}$
and bridge influence is given by $b_i(c_k) = \beta_{NNC}(i) \times k_i^{inter}$.

\subsubsection{Participation Coefficient (PC) \cite{guimera2005functional}} is based on the intra-community and inter-community links distribution. The more the links of a node are distributed across different communities, the higher its participation coefficient. It is defined as follows:
\begin{equation}
\beta_{PC}(i) = 1 - \sum_{c=1}^{N_c}
\left(
\frac{k_{i,c}}{k_i^{tot}}
\right)^2
\end{equation}
where $N_c$ is the total number of communities, $k_{i,c}$ is the number of links node $i$ has in a given community $c$ (can be inter-community or intra-community links), and $k_i^{tot}$ is the total degree of node $i$.

\subsubsection{Community‐based Mediator (CBM) \cite{tulu2018identifying}} takes into consideration the intra-community and inter-community ratio of a node, then it incorporates a random walker and entropy based on the ratio of the different link types. It is defined as follows:
\begin{equation}
\beta_{CBM}(i) = H_i \times \frac{k_i^{tot}}{\sum_{i=1}^{N} k_i}
\end{equation}
where $H_i = [-\sum \rho_i^{intra} log(\rho_i^{intra})] + [- \sum \rho_i^{inter} log(\rho_i^{inter})]$ is the entropy of node $i$ based on its $\rho^{intra}$ and $\rho^{inter}$ which represent the density of the communities a node links to (either its community or external communities), $k_i^{tot}$ is the total degree of node $i$, and $\sum_{i=1}^{N} k_i$ is the total degrees in the network.

\subsubsection{Bridging Centrality (BC) \cite{hwang2006bridging}} extracts node bridges by using betweenness centrality and bridging coefficient. The bridging coefficient quantifies the proximity of a node to high degree nodes. It is defined as follows:
\begin{equation}
\beta_{BC}(i) = \alpha_b(i) \times \mathbb{B}(i)
\end{equation}
where $\alpha_b(i)$ is the classical betweenness centrality of node $i$
and $\mathbb{B}(i)= \frac{k_i^{-1}}{\sum_{j \in \mathcal{N}_1(i)} k_j^{-1}}$ is the bridging coefficient where $\mathcal{N}_1(i)$ is the set of direct neighbors of node $i$.

\section{Datasets and Materials}
\label{sec:mm}
In this section, the 8 real-world online social networks are briefly discussed, alongside the tools applied. Table \ref{tab:table1} reports the basic topological characteristics of the networks. Note that the mixing parameter $\mu$ is defined as the proportion of inter-community links to the total
links in a given network. It is calculated after the community structure is uncovered by the community detection algorithm. 

\subsection{Data}
\subsubsection{FB Ego} this network (ego-facebook) is collected from participants using Facebook. Nodes represent users on Facebook and edges represent online friendships \cite{rossi2015network}.

\subsubsection{FB Princeton} this network (socfb-Princeton12) is collected from Facebook among students at Princeton University. Nodes represent users on Facebook and edges represent online friendships \cite{rossi2015network}.
\subsubsection{FB Caltech} this network (socfb-Caltech36) is collected from the Facebook application among students at Caltech University. Nodes represent users on Facebook and edges represent online friendships \cite{rossi2015network}.

\subsubsection{FB Politician Pages} this network (fb-pages-politician) is collected from Facebook pages. Nodes represent politician pages from different countries created on Facebook and edges represent mutual likes among them \cite{rossi2015network}.

\subsubsection{Retweetes Copenhagen} this network (rt-twitter-copen) is collected from Twitter. Nodes are users on Twitter tweeting in parallel to the United Nations conference in Copenhagen about climate change and edges represent retweets among the users \cite{rossi2015network}.

\subsubsection{DeezerEU} this network (deezer\_europe) is obtained form Deezer, a platform for music streaming. Nodes are Deezer European users and edges represent online friendships \cite{rozemberczki2020characteristic}.

\subsubsection{Hamsterster} this network (petster-friendships-hamster) is obtained from an online social pet network hamsterster.com. Nodes represent users and edges represent friendships among them.
\cite{kunegis2014handbook}.

\subsubsection{PGP} this network (arenas-pgp) is obtained from the web of trust. Nodes are users using the Pretty Good Privacy (PGP) algorithm and edges represent secure information sharing among them \cite{kunegis2014handbook}. 


\begin{table*}[t!]
\centering
\caption{Basic topological properties of the real-world networks. \textit{N} is the total numbers of nodes. E is the number of edges. $<k>$ is the average degree. $<d>$ is the average shortest path. $\nu$ is the density. $\zeta$ is the transitivity (also called global clustering coefficient). $k_{nn}(k)$ is the assortativity (also called degree correlation coefficient). $Q$ is the modularity. $\mu$ is the mixing parameter. * indicates the topological properties of the largest connected component of the network in case it is disconnected.
}
\label{tab:table1}
\begin{tabular}{lcccccccccc}
\hline
Network & $N$ & $E$ & $<k>$ & $<d>$ & $\nu$ & $\zeta$ & $k_{nn}(k)$ & $Q$ & $\mu$\\
\hline

Retweets Copenhagen & 761 & 1,029 & 2.70 & 5.35 & 0.003 & 0.060 & -0.099 & 0.695 &
0.287 \\
FB Caltech* & 762 & 16,651 & 43.70 & 2.23 & 0.057 & 0.291 & -0.066 & 0.389 & 0.410\\
Hamsterster* & 1,788 & 12,476 & 13.49 & 3.45 & 0.007 & 0.090 & -0.088 & 0.391 & 0.298 \\
FB Ego & 4,039 & 88,234 & 43.69 & 3.69 & 0.010 & 0.519 & 0.063 & 0.814 & 0.077\\
FB Politician Pages & 5,908 & 41,729 & 14.12 & 4.66 & 0.002 & 0.301 & 0.018 & 0.836 &
0.111 \\
FB Princeton* & 6,575 & 293,307 & 89.21 & 2.67 & 0.013 & 0.163 & 0.090 & 0.417 & 0.365 \\
PGP & 10,680 & 24,316 & 4.55 & 7.48 & 0.0004 & 0.378 & 0.238 & 0.813 & 0.172 \\
DeezerEU & 28,281 & 92,752 & 6.55 & 6.44 & 0.002 & 0.095 & 0.104 & 0.565 & 0.429\\
\hline
\end{tabular}
\end{table*}
\subsection{Tools}
\subsubsection{Kendall's Tau Correlation} is used to assess the relationship for all possible combinations between classical and community-aware centrality measures. Assume that $R(\alpha)$ and $R(\beta)$ are the ranking lists of a classical centrality and a community-aware centrality, respectively. The correlation value resulted [-1,+1] reveals the degree of ordinal association between the two given sets of ranks. If $R(\alpha_i) > R(\alpha_j)$ and $R(\beta_i) > R(\beta_j)$ or $R(\alpha_i) < R(\alpha_j)$ and $R(\beta_i) < R(\beta_j)$, node pair ($i,j$) is concordant. If $R(\alpha_i) > R(\alpha_j)$ and $R(\beta_i) < R(\beta_j)$ or $R(\alpha_i) < R(\alpha_j)$ and $R(\beta_i) > R(\beta_j)$, node pair ($i,j$) is discordant. If $R(\alpha_i) = R(\alpha_j)$ and/or $R(\beta_i) = R(\beta_j)$, node pair ($i,j$) is neither concordant nor discordant. It is defined as follows:
\begin{equation}
\tau_b(R(\alpha),R(\beta))=\frac{n_{c}-n_{d}}{\sqrt{(n_{c}+n_{disc}+u)(N_{c}+N_{d}+v)}}
\label{eq2}
\end{equation}
where $n_{c}$ and $n_{d}$ stand for the number of concordant and discordant pairs, respectively, and  $u$ and $v$ hold the number of tied pairs in sets $R(\alpha)$ and $R(\beta)$, respectively.

\subsubsection{Rank-Biased Overlap (RBO) \cite{webber2010similarity}}
is capable of placing more emphasis on the top nodes between the two ranked lists $R(\alpha)$ and $R(\beta)$ of classical and community-aware centrality measures. Its value ranges between [0,1]. It is defined as follows:
\begin{equation}
RBO(R(\alpha), R(\beta)) = (1-p){\sum_{d=1}^{\infty} p^{(d-1)}} \frac{|R(\alpha_{d}) \cap
R(\beta_{d})|}{d}
\end{equation}
where $p$ dictates ``user persistence'' and the weight to the top ranks, $d$ is the depth reached on sets $R(\alpha)$ and $R(\beta)$, and $|R(\alpha_{d}) \cap R(\beta_{d})|/{d}$ is the proportion of the similarity overlap at depth $d$. Note that $p$ is set to 0.9 in the experiments.

\subsubsection{Infomap Community Detection Algorithm \cite{rosvall2008maps}} is based on compression of information. The idea is that a random walker on a network is likely to stay longer inside a given community and shorter outside communities. Accordingly, using Huffman coding, each community is defined by a unique codeword and nodes inside communities are defined by other codewords that can be reused in different communities. The optimization algorithm minimizes the coding resulted by the path of the random walker, achieving a concise map of the community structure.

\section{Experimental Results}
\label{sec:res}
\begin{figure*}[t!]
\begin{center}
\includegraphics[width=1\linewidth, height=2.6 in]{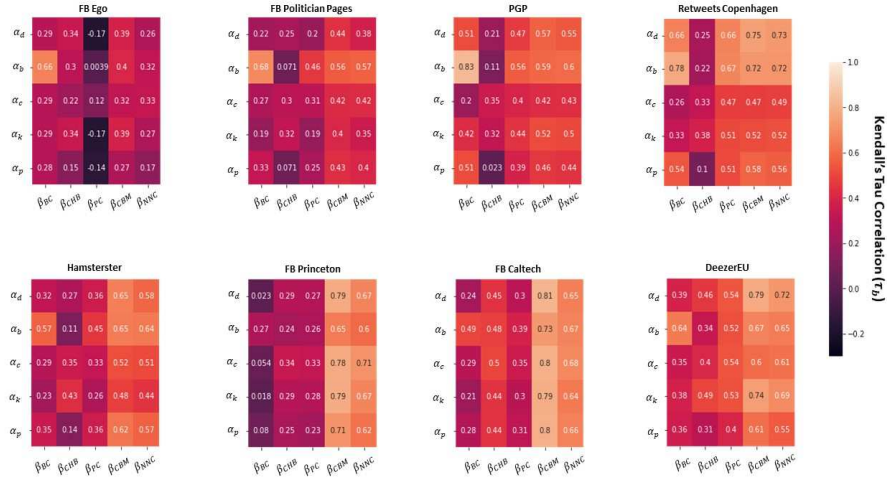}
\end{center}
\caption{{Heatmaps of the Kendall's Tau correlation ($\tau_b$) of real-world networks across the various combinations between classical ($\alpha$) and community-aware ($\beta$) centrality measures. The classical centrality measures are: $\alpha_d$ = Degree, $\alpha_b$ = Betweenness, $\alpha_c$ = Closeness, $\alpha_k$ = Katz, $\alpha_p$ = PageRank. The
community-aware centrality measures are: $\beta_{BC}$ = Bridging centrality, $\beta_{CHB}$ = Community Hub-Bridge, $\beta_{PC}$ = Participation Coefficient, $\beta_{CBM}$ = Community-based Mediator, $\beta_{NNC}$ = Number of Neighboring Communities.}}
\label{fig:corrHeatmapsMain} 
\vspace{-0.4cm}
\end{figure*}

\begin{figure*}[t!]
\begin{center}
\includegraphics[width=1\linewidth, height=2.6 in]{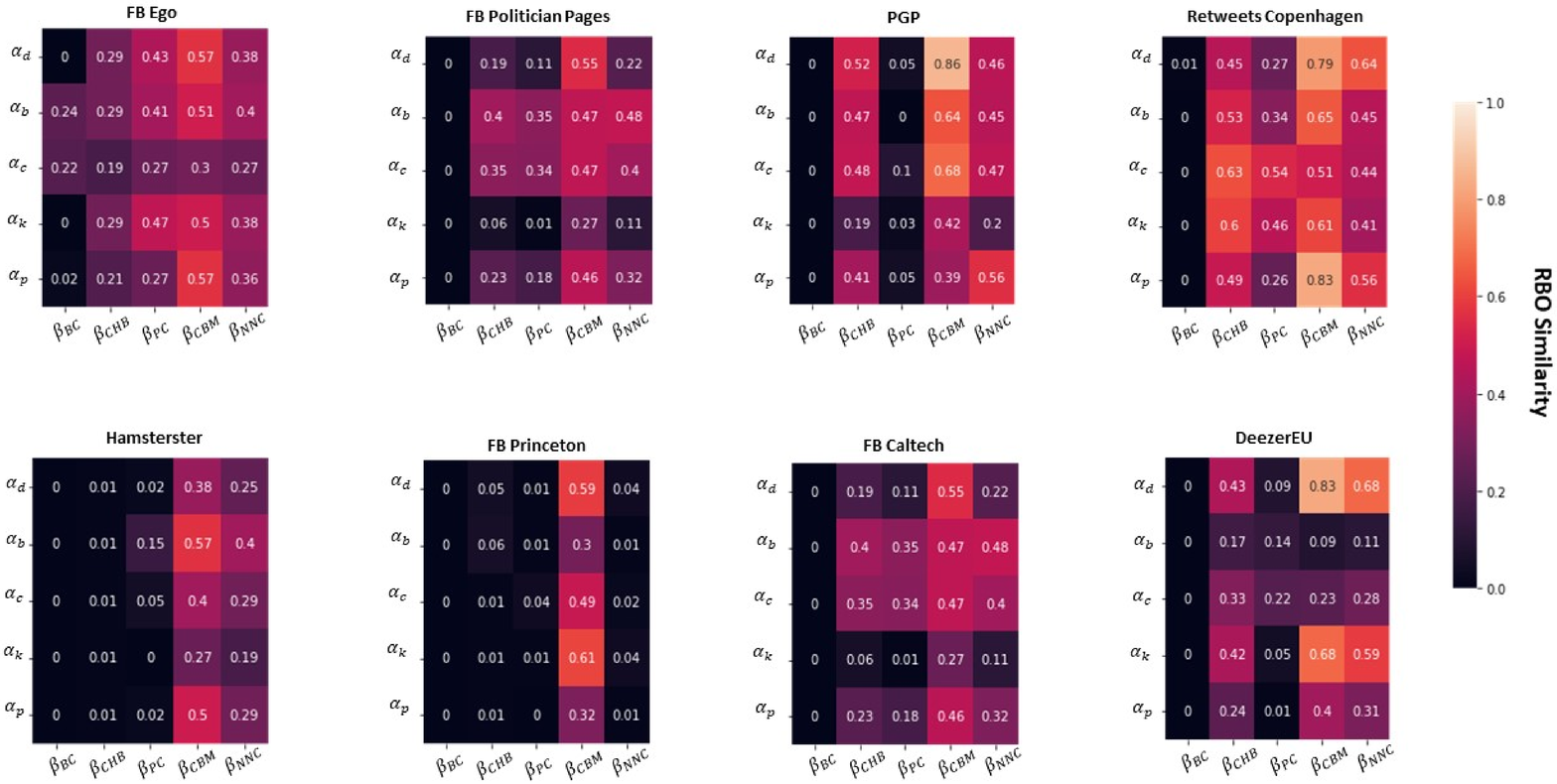}
\end{center}
\caption{{Heatmaps of the RBO similarity at $p$=0.9 of real-world networks across the various combinations between classical ($\alpha$) and community-aware ($\beta$) centrality measures.The classical centrality measures are: $\alpha_d$ = Degree, $\alpha_b$ = Betweenness, $\alpha_c$ = Closeness, $\alpha_k$ = Katz, $\alpha_p$ = PageRank. The
community-aware centrality measures are: $\beta_{BC}$ = Bridging centrality, $\beta_{CHB}$ = Community Hub-Bridge, $\beta_{PC}$ = Participation Coefficient, $\beta_{CBM}$ = Community-based Mediator, $\beta_{NNC}$ = Number of Neighboring Communities.}}
\label{fig:simHeatmapsMain} 
\vspace{-0.4cm}
\end{figure*}
In this section the results of the experiments performed on the real-world networks are reported. The first set of experiments involves calculating Kendall's Tau correlation coefficient for all possible combinations between classical and community-aware centrality measures. The second experiment involves calculating the RBO similarity across all the combinations.

\subsection{Correlation Analysis}
Kendall Tau's correlation is applied on each network given all of the possible combinations between the 5 classical and 5 community-aware centrality measures. The 25 different combinations of the Kendall Tau's correlation for the 8 OSNs are reported in figure \ref{fig:corrHeatmapsMain}. The Kendall's Tau values range from -0.17 to 0.83. Low correlation from -0.17 to 0.3 is characterized by the dark purple color of the heatmaps. Medium correlation from 0.3 to 0.6 is characterized by the fuchsia color. High correlation above 0.6 is characterized by the light pink color.

Networks' heatmaps are arranged from low correlation (FB Ego) to medium-high (DeezerEU) correlation between classical and community-aware centrality measures. Heatmaps show that there are different behaviors among the community-aware centrality measures under study when they are compared to classical centrality measures. Specifically, Bridging centrality
($\beta_{BC}$), Community Hub-Bridge ($\beta_{CHB}$) and Participation Coefficient ($\beta_{PC}$) show consistency in their low correlation with classical centrality measures. On the other hand, Community-based Mediator ($\beta_{CBM}$) and Number of Neighboring Communities ($\beta_{NNC}$) vary across networks. In FB Ego, FB Politician Pages, and PGP, the correlation values are in the low to medium range, while in Hamsterster, FB Princeton, FB Caltech, and DeezerEU they are in the medium to high range.

Note that in Retweets Copenhagen network, the community-aware centrality measures show high correlation with the classical centrality measures degree and betweenness but low to medium correlation with the others. This is with the exception of Community Hub-Bridge ($\beta_{CHB}$) which shows low correlation with all classical centrality measures.

This experiment aims to answer the main research question, that is, do community-aware centrality measures provide distinctive information about the members within OSNs when compared to classical centrality measures? Results show that community-aware centrality measures indeed provide different information from that of classical centrality measures to the members within OSNs. Nonetheless, Bridging centrality ($\beta_{BC}$), Community Hub-Bridge ($\beta_{CHB}$), and Participation Coefficient ($\beta_{PC}$) show consistency in providing distinctive information to the members of 8 networks at hand. They always show low correlation. While Community-based Mediator ($\beta_{CBM}$) and Number of Neighboring Communities ($\beta_{NNC}$) show discrepancy in their behavior from one network to another.

\subsection{Similarity Analysis}
As top nodes are more important than bottom nodes in centrality assessment, RBO is calculated. Moreover, high correlation doesn't necessarily mean high similarity. This is more obvious when ties exist among the rankings of a set. Figure \ref{fig:simHeatmapsMain} shows the RBO similarity heatmaps of the 8 OSNs. The RBO values range from 0 to 0.86. Low similarity from 0 to 0.3 is characterized by the dark purple color. Medium similarity from 0.3 to
0.6 is characterized by the fuchsia color. High similarity over 0.6 is characterized by the light pink color. For comparison purposes, the networks are arranged in the same order as in figure \ref{fig:corrHeatmapsMain}.

Inspecting the heatmaps, Bridging centrality ($\beta_{BC}$) shows almost no similarity with all other classical centrality measures. To a less extent come Community Hub-Bridge ($\beta_{CHB}$) and Participation Coefficient ($\beta_{PC}$) community-aware centrality measures. For these community-aware centralities, the low similarity is consistent across the networks. Community-based Mediator ($\beta_{CBM}$) and Number of Neighboring
Communities ($\beta_{NNC}$) change from one network to another. For example, taking the RBO similarity of the combination ($\alpha_d, \beta_{NNC}$) in DeezerEU, it is equal to 0.68 while in FB Princeton it is equal to 0.04.

This experiment shows consistency with the previous experiment. Indeed, Bridging centrality ($\beta_{BC}$), Community Hub-Bridge ($\beta_{CHB}$), and Participation Coefficient ($\beta_{PC}$) community-aware centrality measures show the lowest similarity to classical centrality measures and their behavior is consistent across the 8 OSNs under study. This case
is similar to the case under Kendall Tau's correlation. However, RBO is more extreme than Kendall's Tau correlation, where low values of similarity can be seen. This is simply due to the RBO definition accounting for ranks. When a group of nodes acquires the same rank, as RBO moves from depth $d$ to $d+1$, the group of tied nodes occurring at $d$ are surpassed and hence account less to the similarity between the two ranked lists.

Referring back to the main research question, indeed, community-aware and classical centrality measures do not convey the same information. Nonetheless, these measures can be divided into two groups. The first group has consistent low similarity with the classical centrality measures while the second group has varying similarity across the networks.

\section{Conclusion}
\label{sec:conc}
Communities have major consequences on the dynamics of a network. Humans tend to form communities within their social presence according to one or many similarity criteria. In addition to that, humans tend to follow other members manifesting power, influence, or popularity, resulting in dense community structures. Centrality measures aim to identify the key members
within OSNs, which is crucial for a lot of strategic applications. However, these measures are agnostic to the community structure. Newly developed centrality measures account for the existence of communities.

Most works have been conducted on classical centrality measures on online social networks. In this work, we shed the light on the relationship between classical and community-aware centrality measures in OSNs. Using 8 real-world OSNs from different platforms, their community structure is uncovered using Infomap. Then, for each network, 5 classical and 5 community-aware centrality measures are calculated. After that, correlation and similarity evaluation between all possible classical and community-aware centrality measures is conducted. Results show that globally these two types of centrality do not convey the same information. Moreover,
community-aware centrality measures exhibit two behaviors. The first set (Bridging centrality, Community Hub-Bridge, and Participation Coefficient) exhibit low correlation and low similarity for all the networks under study. The second set (Community-based Mediator and Number of Neighboring
Communities) shows varying correlation and similarity across networks.

Results of this study suggest that community-aware centrality measures are worth looking into when searching for key members in OSNs, as they provide different information from classical centrality measures. This work opens future research directions. Further study will investigate the effect of network topology on the relationship between classical and community-aware
centrality measures and whether results are consistent using different community detection algorithms.

%
%

\bibliographystyle{unsrt}
\bibliography{bibtech}

\end{document}